\documentclass{article}

\usepackage{amsmath}
\usepackage{bm} 
\usepackage{amssymb}
\usepackage{booktabs}

\usepackage{amsmath,amsthm,amsfonts}
\usepackage{graphicx}
\usepackage{booktabs}
\usepackage{parskip}

\renewenvironment{abstract}{\vskip.075in\centerline{\large\bf
Abstract}\vspace{0.5ex}\begin{quote}}{\par\end{quote}\vskip 1ex}

\usepackage{natbib}

\title{Generalised Wishart Processes}

\author{\begin{tabular}{ccc}
    {Andrew Gordon Wilson\thanks{http:\///mlg.eng.cam.ac.uk\//andrew}} &\qquad \qquad& {Zoubin Ghahramani}\\
   Department of Engineering && Department of Engineering\\ 
   University of Cambridge && University of Cambridge\\
   {\tt agw38@cam.ac.uk} && {\tt zoubin@eng.cam.ac.uk}
  \end{tabular}
}

\date{}

\begin{document}

\maketitle

\begin{abstract}
We introduce a stochastic process with Wishart marginals: the \textit{generalised Wishart process} (GWP).
It is a collection of positive semi-definite random matrices indexed by any arbitrary dependent variable.
We use it to model dynamic (e.g.\ time varying) covariance matrices $\Sigma(t)$.  Unlike existing models,
it can capture a diverse class of covariance structures, it can easily handle missing data, the dependent 
variable can readily include covariates other than time, and it scales well with
dimension; there is no need for free parameters, and optional parameters are easy to interpret. We
describe how to construct the GWP, introduce general procedures for inference and predictions, and 
show that it outperforms its main competitor, multivariate GARCH, even on financial data that especially
suits GARCH.  We also show how to predict the mean $\bm{\mu}(t)$ of a multivariate
process while accounting for dynamic correlations.
\end{abstract}

\section{Introduction}

Imagine the price of the NASDAQ composite index increased dramatically today.
Will it continue to rise tomorrow?  Should you invest?  Perhaps this is the beginning 
of a trend, but it may also be an anomaly.  Now suppose you discover that every major 
equity index -- FTSE, NIKKEI, TSE, etc. -- has also risen.  
Instinctively the rise in NASDAQ was not anomalous, because this market is correlated
with other major indices. This is an example of how multivariate models which account for
correlations can be better than univariate models at making univariate 
predictions.

In this paper, we are concerned with modelling the dynamic covariance matrix $\Sigma(t)$ 
of high dimensional data sets (multivariate volatility).  These models are especially 
important in econometrics.  \citet{engle2009} remark that ``The price of essentially every derivative 
security is affected by swings in volatility. Risk management models used by financial institutions 
and required by regulators take time-varying volatility as a key input. Poor appraisal of the risks to come can 
leave investors excessively exposed to market fluctuations or institutions hanging on a precipice of inadequate capital''.  
Indeed, Robert Engle and Clive Granger won the 2003 Nobel prize in economics ``for methods of analysing 
economic time series with time-varying volatility''. The returns on major equity indices and currency
exchanges are thought to have a time changing variance and zero mean, and GARCH \citep{bollerslev86}, 
a generalization of Engle's ARCH \citep{engle82}, is arguably unsurpassed at predicting the 
volatilities of returns on these equity indices and currency exchanges 
\citep{granger05, hansen05, engle2009}.  Multivariate volatility models can be used 
to understand the dynamic correlations (or \textit{co-movement}) 
between equity indices, and can make better univariate predictions than univariate models.  
A good estimate of the covariance matrix $\Sigma(t)$ is also necessary for portfolio
management.  An optimal portfolio allocation $\bm{w}^*$ is said to maximise
the Sharpe ratio \citep{sharpe1966}:
\begin{equation}
\frac{\text{Portfolio return}}{\text{Portfolio volatility}} = \frac{\bm{w}^{\top}\bm{r}(t)}{\sqrt{\bm{w}^{\top}\Sigma(t)\bm{w}}} \,, 
\end{equation}
where $\bm{r}(t)$ are expected returns for each asset and $\Sigma(t)$ is the predicted covariance matrix for these returns.  One 
may also wish to maximise the portfolio return $\bm{w}^{\top}\bm{r}(t)$ for a fixed level of volatility:
$\sqrt{\bm{w}^{\top}\Sigma(t)\bm{w}} = \lambda$.  Sharpe, Markowitz and Merton jointly received a Nobel prize for
portfolio theory \citep{markowitz1952, merton1972}.  Multivariate volatility models 
are also used to understand \textit{contagion}: the transmission of a financial shock from one entity to another \citep{bae2003}.  And generally -- in econometrics,
machine learning, climate science, or otherwise -- it is useful to know the dynamic correlations between multiple entities.

Despite their importance, existing multivariate volatility models suffer from tractability
issues and a lack of generality.  For example, multivariate GARCH (MGARCH) has a number of free 
parameters that scales with dimension to the fourth power, and interpretation and estimation of these
parameters is difficult to impossible \citep{anna09, gourieroux97}, given the constraint that $\Sigma(t)$ must
be positive definite at all points in time.  Thus MGARCH, and alternative multivariate stochastic 
volatility (MSV) models\footnote{MSV models, pioneered by \citet{harvey94}, assume volatility follows a random process, unlike GARCH which assumes it is
a deterministic function of the past.}, are generally limited to studying processes with less than 5 components \citep{gourieroux09}.
Recent efforts have led to simpler but less general models, which make assumptions such as 
constant correlations \citep{bollerslev90} -- leaving only the diagonal entries of $\Sigma(t)$ to vary.  

We hope to unite machine learning and econometrics in an effort to solve these problems.  We
introduce a stochastic process with Wishart marginals: the \textit{generalised Wishart process} (GWP).  
It is a collection of positive semi-definite random matrices indexed by any arbitrary dependent variable $\bm{z}$.
We call it the \textit{generalised} Wishart process, since it is a generalisation of the first Wishart process 
defined by \citet{bru91}.  Bru's Wishart process has recently been used \citep{gourieroux09} in
multivariate stochastic volatility (MSV) models \citep{philipov06, harvey94}.  This prior work on Wishart processes 
is limited for several reasons: 1) it assumes the dependent variable is a scalar, 2) it is restricted to using an 
Ornstein-Uhlenbeck covariance structure\footnote{An \textit{Ornstein-Uhlenbeck process} \citep{uhlenbeck30} was first introduced
to model the velocity of a particle undergoing Brownian motion.} (which means $\Sigma(t+a)$ and $\Sigma(t-a)$ are independent given $\Sigma(t)$, and complex
dependencies cannot be captured), 3) it is autoregressive, and 4) there are no general learning and inference procedures.  
The generalised Wishart process (GWP) addresses all of these issues.  Specifically, in the GWP formulation,
\begin{itemize}
 \item The dependent variable can come from any arbitrary index set, just as easily as it can represent time.  This allows one to effortlessly
       condition on covariates like interest rates.
 \item One can easily handle missing data.
 \item One can easily specify a range of covariance structures (periodic, smooth, Ornstein-Uhlenbeck, \dots).
 \item We develop Bayesian inference procedures to make predictions, and to learn distributions over any relevant parameters.  Aspects of the covariance
       structure are learned from data, rather than being a fixed property of the model.
\end{itemize}
Overall, the GWP is versatile and simple.  It does not require any free parameters, and any optional parameters are easy to
interpret. For this reason, it also scales well with dimension.  Yet, the GWP provides an especially 
general description of multivariate volatility -- more so than the most general MGARCH specifications.  In the next section, 
we review Gaussian processes (GPs), which are used to construct the Wishart process.  In the following sections we then review the Wishart distribution, present the 
GWP construction, introduce procedures for inference and predictions, review the main competitor, MGARCH, and present experiments 
that show how the GWP outperforms MGARCH on simulated and financial data.  These experiments include a 5 dimensional data set, 
based on returns for NASDAQ, FTSE, NIKKEI, TSE, and the Dow Jones Composite, and a set of returns for 3 foreign currency exchanges.  
In a subsequent version we will also present a 200 dimensional experiment to show how the GWP can be used to study high dimensional problems.   

Also, although it is not the focus of this paper, we show in the inference section how the GWP can additionally be used as 
part of a new GP based regression model that accounts for \textit{changing} correlations.  In other words,
it can be used to predict the mean $\bm{\mu}(t)$ together with the covariance matrix $\Sigma(t)$ of a 
multivariate process.  Alternative GP based multivariate regression models for $\bm{\mu}(t)$, which account for
fixed correlations, were recently introduced by \citet{bonilla2008}, \citet{teh2005}, and \citet{boyle2004}.

\section{Gaussian Processes}
We briefly review Gaussian processes, since the generalised Wishart process is
constructed from GPs.  For more detail, see \citet{rasmussen06}. 

A Gaussian process is a collection of random variables, any finite number of which have a joint Gaussian 
distribution.  Using a Gaussian process, we can define a distribution over functions $u(\bm{z})$:
\begin{equation}
u(\bm{z}) \sim \mathcal{GP}(m(\bm{z}),k(\bm{z},\bm{z}')) \,,
\end{equation}
where $\bm{z}$ is an arbitrary (potentially vector valued) dependent variable, and
the mean $m(\bm{z})$ and kernel function $k(\bm{z},\bm{z}')$ are respectively defined as 
\begin{align}
m(\bm{z}) &= \mathbb{E}[u(\bm{z})] \,, \\
k(\bm{z},\bm{z}') &=  \text{cov}(u(\bm{z}),u(\bm{z}'))\,.
\end{align}
This means that any collection of function values has a joint Gaussian distribution:
\begin{equation}
(u(\bm{z}_1),u(\bm{z}_2),\dots,u(\bm{z}_N))^{\top} \sim \mathcal{N}(\bm{\mu},K) \,,
\end{equation}
where the $N \times N$ Gram matrix $K$ has entries $K_{ij} = k(\bm{z}_i,\bm{z}_j)$, and
the mean $\bm{\mu}$ has entries $\bm{\mu}_i = m(\bm{z}_i)$.  The properties of these
functions (smoothness, periodicity, etc.) are determined by the kernel function.  The
squared exponential kernel is popular:
\begin{equation}
 k(\bm{z},\bm{z}') = \exp(-0.5||\bm{z}-\bm{z}'||^2/l^2) \,.
\end{equation}
Functions drawn from a Gaussian process with this kernel function are smooth, and
can display long range trends.  The length-scale \textit{hyperparameter} $l$ is easy
to interpret: it determines how much the function values $u(\bm{z})$ and
$u(\bm{z}+\bm{a})$ depend on one another, for some constant $\bm{a}$.  

The autoregressive process 
\begin{align}
u(t+1) = u(t) + \epsilon(t) \,, \\
\epsilon(t) \sim \mathcal{N}(0,1) \,,
\end{align}
is an example of a Gaussian process with a fixed covariance structure. 

\section{Wishart Distribution}
The Wishart distribution defines a probability density function over positive definite matrices $S$:
\begin{equation}
 p(S|V,\nu) = \frac{|S|^{(\nu - D - 1)/2}}{2^{\nu D/2}|V|^{\nu/2}\Gamma_D(\nu/2)}\exp(-\frac{1}{2}\text{tr}(V^{-1}S))\,,
\end{equation}
where $V$ is a $D \times D$ positive definite scale matrix, and $\nu > 0$ is the number of degrees of freedom.
This distribution has mean $\nu V$ and mode $(D-\nu-1)V$ for $\nu \geq D+1$. 
$\Gamma_D(\cdot)$ is the multivariate gamma function:
\begin{equation}
 \Gamma_D(\nu /2) = \pi^{D(D-1)/4} \prod_{j=1}^D \Gamma(\nu/2 + (1-j)/2)\,. 
\end{equation}
The Wishart distribution is a multivariate generalisation of the Gamma distribution when $\nu$ is real valued,
and the chi-square ($\chi^2$) distribution when $\nu$ is integer valued.  The sum of squares of univariate
Gaussian random variables is chi-squared distributed.  Likewise, the sum of outer products of multivariate
Gaussian random variables is Wishart distributed:
\begin{equation}
 S = \sum_{i=1}^{\nu} \bm{u}_i \bm{u}_i^{\top} \sim \mathcal{W}_D(V,\nu)\,,
\end{equation}
where the $\bm{u}_i$ are i.i.d. $\mathcal{N}(\bm{0},V)$ $D$-dimensional random variables, and  
$\mathcal{W}_D(V,\nu)$ is a Wishart distribution with $D \times D$ scale matrix $V$, and $\nu$ degrees of freedom.  
$S$ is a $D \times D$ positive definite matrix.  
If $D=V=1$ then $\mathcal{W}$ is a chi-square distribution with $\nu$ degrees of freedom. $S^{-1}$ has
the inverse Wishart distribution, $\mathcal{W}_D^{-1}(V^{-1},\nu)$, which is a conjugate prior for covariance matrices 
of zero mean Gaussian distributions.  This means that for data $\mathcal{D}$ if a prior $p(R)$ is inverse Wishart, and the 
likelihood $p(\mathcal{D}|R)$ is Gaussian with zero mean, then the posterior 
$p(R|\mathcal{D})$ is also inverse Wishart.

\section{Generalised Wishart Process Construction}

We saw that the Wishart distribution is constructed from multivariate Gaussian distributions.  Essentially,
by replacing these Gaussian distributions with Gaussian processes, we define a process with Wishart marginals
-- the \textit{generalised Wishart process}.  It is a collection of positive semi-definite random matrices
indexed by any arbitrary (potentially high dimensional) dependent variable $\bm{z}$.  For clarity, we assume that 
time is the dependent variable, even though it takes no more effort to use a vector-valued variable $\bm{z}$ 
from any arbitrary set.  Everything we write would still apply if we replaced $t$ with $\bm{z}$.

Suppose we have $\nu D$ independent Gaussian process functions, $u_{id}(t) \sim \mathcal{GP}(0,k)$, 
where $i=1,\dots,\nu$ and $d=1,\dots,D$.  This means $\text{cov}(u_{id}(t),u_{id}(t')) = k(t,t')\delta_{ii'}\delta_{dd'}$, 
and $(u_{id}(t_1), u_{id}(t_2), \dots, u_{id}(t_N))^{\top} \sim \mathcal{N}(0,K)$, where $\delta_{ij}$ is
the Kronecker delta, and $K$ is an $N \times N$ Gram matrix with elements $K_{ij} = k(t_i,t_j)$.  
Let $\hat{\bm{u}}_i(t) = (u_{i1}(t),\dots,u_{iD}(t))^{\top}$, 
and let $L$ be the lower Cholesky decomposition of a $D \times D$ scale matrix $V$, such that $LL^{\top} = V$.
Then at each $t$ the covariance matrix $\Sigma(t)$ has a Wishart marginal distribution,
\begin{equation}
\Sigma(t) =  \sum_{i=1}^{\nu} L \hat{\bm{u}}_i(t) \hat{\bm{u}}_i^{\top}(t) L^{\top} \sim \mathcal{W}_D(V,\nu)\,,  \label{eqn: sigmagen}
\end{equation}
subject to the constraint that the kernel function $k(t,t) = 1$.  

We can understand \eqref{eqn: sigmagen} as follows.  Each element of the vector $\hat{\bm{u}}_i(t)$ is a univariate Gaussian with zero mean and variance
$k(t,t)=1$.  Since these elements are uncorrelated, $\hat{\bm{u}}_i(t) \sim \mathcal{N}(\bm{0},I)$.  Therefore
$L\hat{\bm{u}}_i(t) \sim \mathcal{N}(0,V)$, since $\mathbb{E}[L\hat{\bm{u}}_i(t)\hat{\bm{u}}_i(t)^{\top}L^{\top}] = LIL^{\top} = LL^{\top} = V$.
We are summing the outer products of $\mathcal{N}(0,V)$ random variables, and there are $\nu$ terms in the sum, so 
by definition this has a Wishart distribution $\mathcal{W}_{D}(V,\nu)$.
\begin{figure}
\centering
\includegraphics[scale=.4]{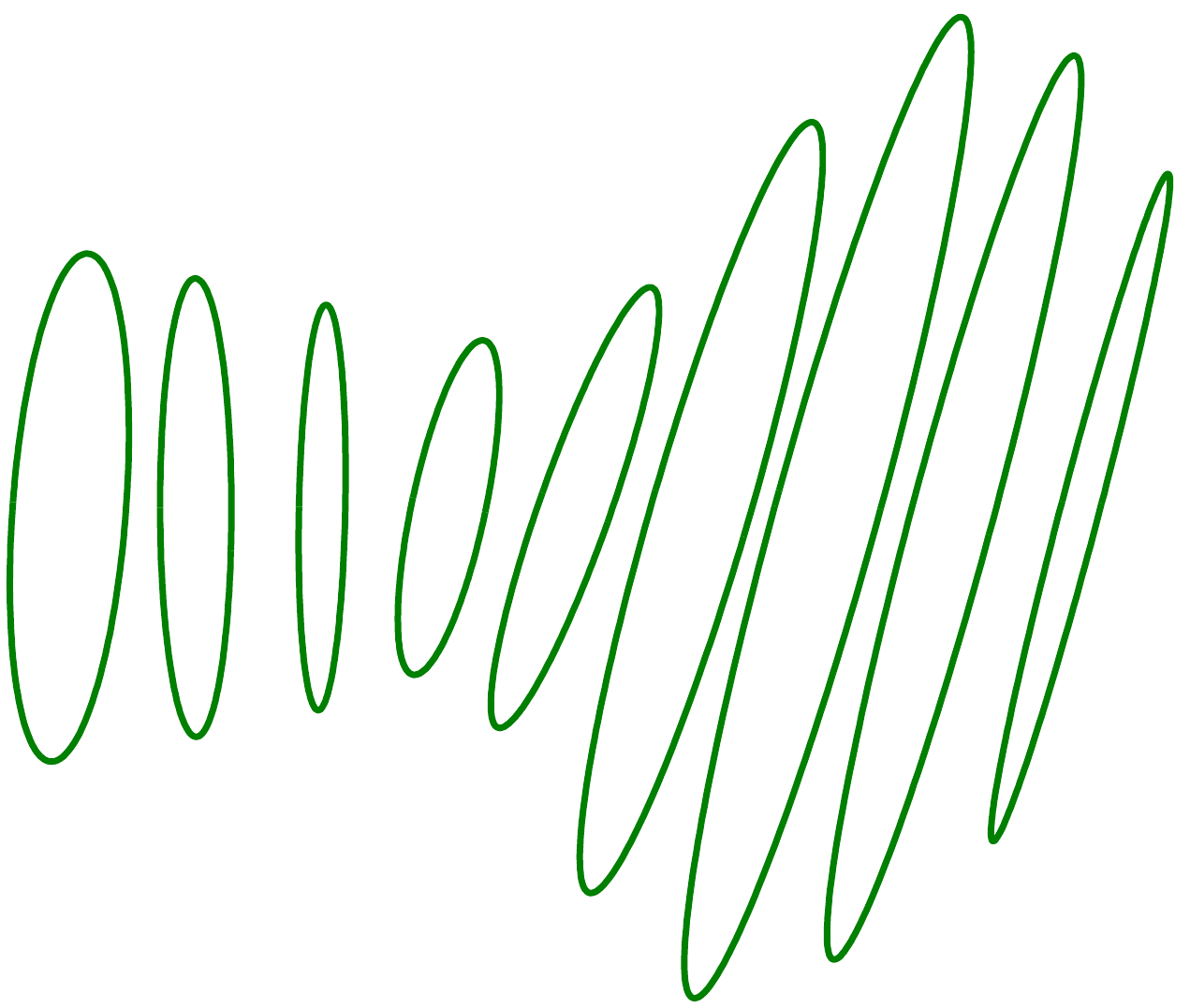}
\caption[Draw from a Wishart process prior.]
{\small A draw from a generalised Wishart process (GWP).  Each ellipse is a $2 \times 2$ covariance matrix indexed by time, 
which increases from left to right.  The rotation indicates the correlation between the two variables, and the major and 
minor axes scale with the eigenvalues of the matrix.  Like a draw from a Gaussian process is a collection of function values
indexed by time, a draw from a GWP is a collection of matrices indexed by time.}
\label{fig: ellipses}
\end{figure}

We write $\Sigma(t) \sim \mathcal{GWP}(V,\nu,k(t,t'))$ to mean that $\Sigma(t)$ is a collection of positive semi-definite 
random matrices with $\mathcal{W}_D(V,\nu)$ marginal distributions.  Assuming the dependent variable is time, a draw from a Wishart
process is a collection of matrices indexed by time (Figure \ref{fig: ellipses}), much like a draw from a Gaussian process is a collection of function values indexed by time.  

Using this construction, we can also define a generalised \textit{inverse} Wishart process (GIWP).  If $\Sigma(t) \sim \mathcal{GWP}$,
then inversion at each value of $t$ defines a draw $R(t) = \Sigma(t)^{-1}$ from the GIWP.  The conjugacy of the GIWP 
with a Gaussian likelihood could be useful when doing Bayesian inference.  

We can further extend this construction by replacing the Gaussian processes with \textit{copula processes} \citep{wilson2010}.  
For example, as part of Bayesian inference we could learn a mapping that would transform the Gaussian processes $u_{id}$ to 
Gaussian copula processes with marginals that better suit the covariance structure of our data set; the
result is a \textit{Wishart copula process}. 

The formulation we outlined in this section is different from other multivariate volatility models in that one can specify a kernel 
function $k(t,t')$ that controls how $\Sigma(t)$ varies with $t$ -- for example, $k(t,t')$ could be periodic -- and $t$ need not be time:
it can be an arbitrary dependent variable, including covariates like interest rates.  In the next section we introduce, 
for the first time, general inference procedures for making predictions when using a Wishart process prior.  These are based on recently 
developed Markov chain Monte Carlo techniques \citep{murray10}.  We also introduce a new method for doing multivariate GP based regression with
\textit{dynamic} correlations.

\section{Bayesian Inference}

\begin{figure}
\centering
\includegraphics[scale=.3]{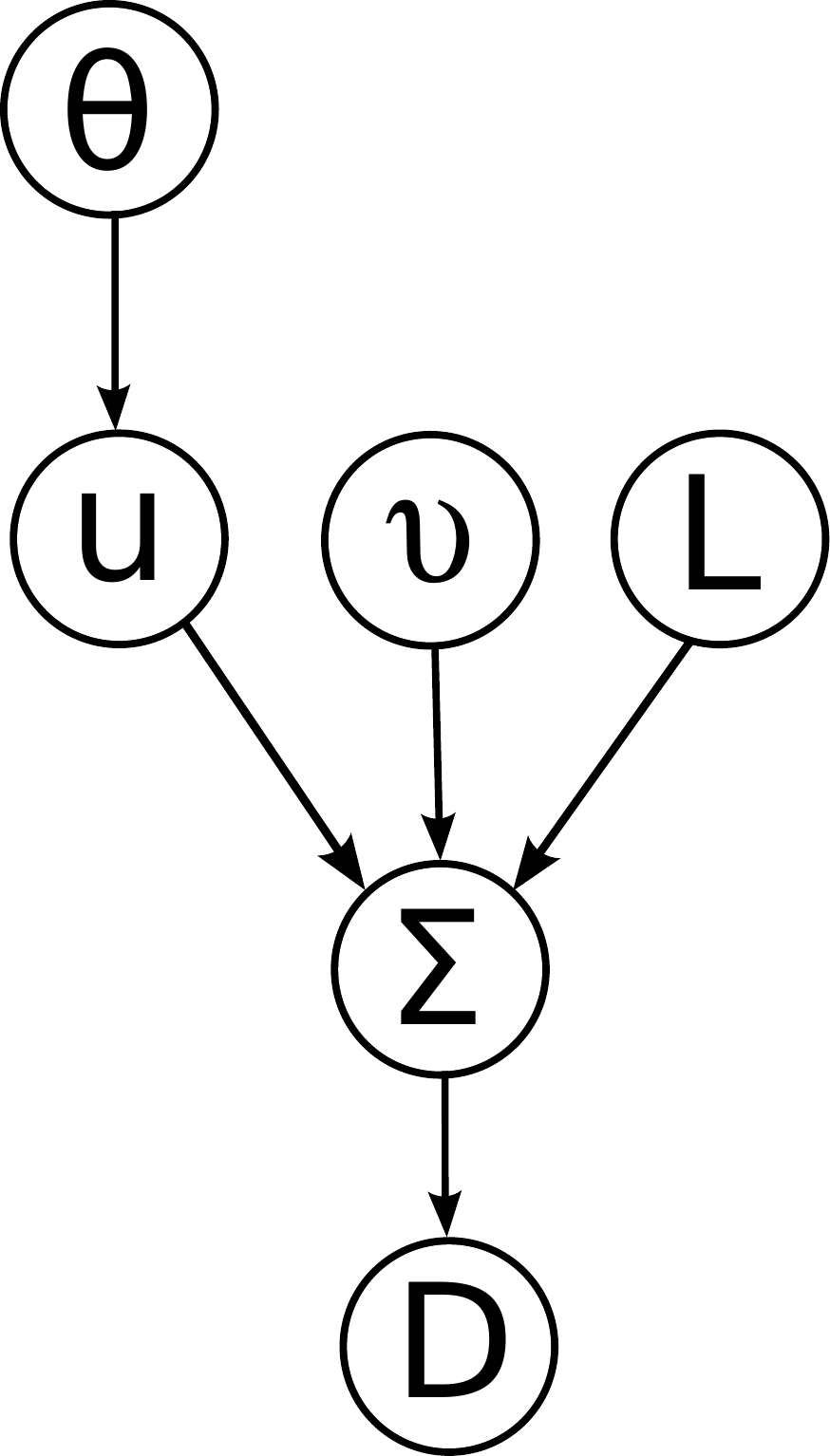}
\caption[Graphical model of the generalised Wishart process]
{\small Graphical model of the generalised Wishart process. $\bm{u}$ is a vector of GP function values, $\bm{\theta}$ are GP hyperparameters,
$L$ is the lower Cholesky decomposition of the scale matrix ($LL^{\top}=V$), $\nu$ are the degrees of freedom, and $\Sigma$ is the covariance matrix.}
\label{fig: gmodel}
\end{figure}

Assume we have a generalised Wishart process prior on a dynamic $D \times D$ covariance matrix:
\begin{equation}
 \Sigma(t) \sim \mathcal{GWP}(V,\nu,k).
\end{equation}
We want to infer the posterior $\Sigma(t)$ given a $D$-dimensional data set $\mathcal{D} = \{\bm{x}(t_n) : n = 1,\dots,N\}$.
We explain how to do this for a general likelihood function, $p(\mathcal{D}|\Sigma(t))$, by finding the 
posterior distributions over the parameters in the model, given the data $\mathcal{D}$. 
These parameters are: a vector of all relevant GP function values $\bm{u}$, the hyperparameters of the 
GP kernel function $\bm{\theta}$, the degrees of freedom $\nu$, and $L$, the lower cholesky decomposition 
of the scale matrix $V$ ($LL^{\top} = V$). The graphical model in Figure \ref{fig: gmodel} shows all the relevant parameters
and conditional dependence relationships.
 
We can sample from these posterior distributions using Gibbs sampling \citep{geman1984}, a Markov chain Monte Carlo algorithm where initialising 
$\{\bm{u},\bm{\theta},L,\nu\}$ and then sampling in cycles from
\begin{align}
 p(\bm{u}|\bm{\theta},L,\nu,\mathcal{D}) &\propto p(\mathcal{D}|\bm{u},L,\nu)p(\bm{u}|\bm{\theta}) \,, \label{eqn: up} \\
 p(\bm{\theta}|\bm{u},L,\nu,\mathcal{D}) &\propto p(\bm{u}|\bm{\theta})p(\bm{\theta})\,, \label{eqn: tp} \\
 p(L|\bm{\theta},\bm{u},\nu,\mathcal{D}) &\propto p(\mathcal{D}|\bm{u},L,\nu)p(L)\,, \label{eqn: lchol} \\
 p(\nu|\bm{\theta},\bm{u},L,\mathcal{D}) &\propto p(\mathcal{D}|\bm{u},L,\nu)p(\nu)\,, \label{eqn: nup}
\end{align}
will converge to samples from $p(\bm{u},\bm{\theta},L,\nu|\mathcal{D})$.  We will successively 
describe how to sample from the posterior distributions \eqref{eqn: up}, \eqref{eqn: tp}, \eqref{eqn: lchol}, and \eqref{eqn: nup}.  
In our discussion we assume there are $N$ data points (one at each time step or input), and $D$ dimensions.  We then explain how
to make predictions of $\Sigma(t_*)$ at some test input $t_*$.  Finally, we discuss a potential likelihood function, and how the
GWP could also be used as a new GP based model for multivariate regression with outputs that have changing correlations.

\subsection{Sampling the GP functions}
In this section we describe how to sample from the posterior distribution \eqref{eqn: up} over the Gaussian process function values $\bm{u}$.
We order the entries of $\bm{u}$ by fixing the degrees of freedom and dimension, and running the time steps from $n=1,\dots,N$.  
We then increment dimensions, and finally, degrees of freedom.  So $\bm{u}$ is a vector of length $N D \nu$.  As before, let
$K$ be an $N \times N$ Gram matrix, formed by evaluating the kernel function at all pairs of training inputs.  Then the 
prior $p(\bm{u}|\bm{\theta})$ is a Gaussian distribution with $N D \nu \times N D \nu$ block diagonal covariance matrix $K_B$, formed 
using $D \nu$ of the $K$ matrices; if the hyperparameters of the kernel function change depending on dimension or degrees 
of freedom, then these $K$ matrices will be different from one another.  In short,
\begin{equation}
 p(\bm{u}|\bm{\theta}) = \mathcal{N}(0,K_B)\,.
\end{equation}
With this prior, and the likelihood formulated in terms of the other parameters, we can sample from the posterior \eqref{eqn: up}.
Sampling from this posterior is difficult, because the Gaussian process function values are highly correlated by the $K_B$ matrix.
We use Elliptical Slice Sampling \citep{murray10}: it has no free parameters, jointly updates every element of $\bm{u}$, and was
especially designed to sample from posteriors with correlated Gaussian priors.  We found it effective.

\subsection{Sampling the other parameters}
We can similarly obtain distributions over the other parameters. The priors we use will depend on the data we 
are modelling.  We placed a vague lognormal prior on $\bm{\theta}$ and sampled from the posterior \eqref{eqn: tp} 
using axis aligned slice sampling if $\bm{\theta}$ was one dimensional, and Metropolis Hastings otherwise.  We also used 
Metropolis Hastings to sample from \eqref{eqn: lchol}, with a spherical Gaussian prior on the elements of $L$.  To sample 
\eqref{eqn: nup}, one can use reversible jump MCMC \citep{green1995, robert2004}.  But in our experiments we set $\nu = D+1$, 
and found it effective.  Although learning $L$ is not expensive, one might simply wish to set it by taking
the empirical covariance of the data set, dividing by the degrees of freedom, and then taking the lower
cholesky decomposition.

\subsection{Making predictions}

Once we have learned the parameters $\{\bm{u},\bm{\theta},L,\nu\}$, we can find a distribution over
$\Sigma(t_*)$ at a test input $t_*$.  To do this, we must infer the distribution over $\bm{u}_*$ -- 
all the relevant GP function values at $t_*$:
\begin{align}
 \bm{u}_* = [&u_{11}(t_*) \dots u_{1D}(t_*) u_{21}(t_*) \dots u_{2D}(t_*)  \\
            &\dots u_{\nu 1}(t_*) \dots u_{\nu D}(t_*)]^{\top} \,. \notag
\end{align}
Consider the joint distribution over $\bm{u}$ and $\bm{u}_*$:
\begin{align}
\begin{bmatrix} \bm{u} \\ \bm{u}_* \end{bmatrix} 
\sim \mathcal{N}(\bm{0},
\left[
\begin {array}{cc}
K_B & A^{\top}\\
\noalign{\medskip}
A & I_p\
\end {array}
\right])\,.
\end{align}
Supposing that $\bm{u}_*$ and $\bm{u}$ respectively have $p$ and $q$ elements, then $A$ is a $p \times q$ 
matrix of covariances between the GP function values $\bm{u}_*$ and $\bm{u}$ at all 
pairs of the training and test inputs:  $A_{ij} = k_i(t_*,t_{\text{mod}(N+1,j)})$ if $1+(i-1)N \leq j \leq iN$, and
$0$ otherwise.  The kernel function $k_i$ may differ from row to row, if it changes depending on the 
degree of freedom or dimension; for instance, we could have a different
length-scale for each new dimension.  $I_p$ is a $p \times p$ identity matrix 
representing the prior independence between the GP function values in $\bm{u}_*$.  Conditioning on $\bm{u}$, we find
\begin{equation}
\bm{u}_* | \bm{u} \sim \mathcal{N}(A K_B^{-1} \bm{u}, I_p - AK_B^{-1}A^{\top}) \,.
\end{equation}
We can then construct $\Sigma(t_*)$ using equation \eqref{eqn: sigmagen}
and the elements of $\bm{u}_*$.

\subsection{Likelihood function}
So far we have avoided making the likelihood explicit; the inference procedure
we described will work with a variety of likelihoods parametrized through a
matrix $\Sigma(t)$, such as the multivariate $t$ distribution.  However, assuming
for simplicity that each of the variables $\bm{x}(t_n)$ has a Gaussian distribution,
\begin{equation}
\bm{x}(t) \sim \mathcal{N}(\bm{\mu}(t),\Sigma(t)), \label{eqn: xsigma}
\end{equation}
then the likelihood is
\begin{align}
 p(\mathcal{D}|\bm{\mu}(t),\Sigma(t)) = \prod_{n=1}^{N} p(\bm{x}(t_n)|\bm{\mu}(t_n),\Sigma(t_n)) \notag \\ 
= \prod_{n=1}^N |2\pi \Sigma(t_n)|^{-1/2} \exp [-\frac{1}{2} \bm{w}(t_n)^{\top}\Sigma(t_n)^{-1}\bm{w}(t_n)] \label{eqn: likeliwish},
\end{align}
where $\bm{w}(t_n) = \bm{x}(t_n) - \bm{\mu}(t_n)$.  We can learn a distribution over $\bm{\mu}(t)$, 
in addition to $\Sigma(t)$. Here are three possible specifications of $\bm{\mu}$:
\begin{align}
 \bm{\mu}(t) &= \sum_{i=1}^{\nu} L\hat{\bm{u}}_i(t) \,, \label{eqn: mugen1} \\
 \bm{\mu}(t) &= \sum_{i=1}^{\nu} L\hat{\bm{u}}_i(t) + \hat{\bm{u}}_{\nu + 1}(t) \,, \label{eqn: mugen2}  \\
 \bm{\mu}(t) &= \hat{\bm{u}}_{\nu + 1}(t) \,. \label{eqn: mugen3} 
\end{align}
In \eqref{eqn: mugen1}, the mean function is directly coupled to the covariance matrix in \eqref{eqn: sigmagen},
since they are both constructed using the same Gaussian processes.  As the components of $\bm{\mu}$ increase in 
magnitude, so do the entries in $\Sigma$.  This is a desirable property if we expect, for example, high returns 
to be associated with high volatility.  This property is encouraged but not enforced in \eqref{eqn: mugen2}, where a 
separate vector of Gaussian processes $\bm{u}_{\nu +1}(t)$ is introduced into the expression for the mean function, 
but not the expression for $\Sigma(t)$.  In \eqref{eqn: mugen3}, the mean function is solely this separate vector 
of Gaussian processes.  In each of these cases, we can make mean predictions by inferring distributions over the 
GP function values $\bm{u}$, as outlined above.  And so in each case, the GWP is being used as a GP based regression 
model which accounts for multiple outputs that have \textit{changing} correlations. 

Alternative models, which account for fixed correlations, have recently
been introduced by \citet{bonilla2008}, \citet{teh2005}, and \citet{boyle2004}. Rather than use a GP based
regression, as in \eqref{eqn: mugen1}-\eqref{eqn: mugen3}, \citet{gelfand2004} combine a spatial Wishart process 
with a parametric linear regression on the mean, to make correlated mean predictions in a spatial setting.  
Generally their methodology is substantially different: the correlation structure is a fixed property of their 
method (they do not learn the parameters of a kernel function), and they are not developing a multivariate volatility 
model, so are not interested in explicitly learning or evaluating the accuracy of the dynamic correlations; in fact,
they do not explain how to make predictions of $\Sigma$ at a test input.  Further, they do not sample
$\bm{\theta}, L,$ or $\nu$, and their inference is not explained except that their sampling relies solely on Metropolis Hastings with 
Gaussian proposals, which will not scale to high dimensions, and will not mix efficiently as the strong GP prior correlations are 
not accounted for. They fix $L$ as diagonal, which significantly limits the correlation structure of the dynamic covariance
matrices (e.g. $\Sigma(t)$), as does fixing $\bm{\theta}$, which we have empirically found to severely affect the quality
of predictions.  In this paper we focus on making predictions of $\Sigma(t)$, setting $\bm{\mu}=0$.

\subsection{Computational complexity}

In contrast to the alternatives, our method scales exceptionally nicely with dimension. MGARCH, the 
main competitor, is limited to 5 dimensions, at which point severe assumptions -- such as constant correlations 
-- are needed for tractability in higher dimensions.  We conjecture that other Wishart process models are similarly
limited, with \citet{gelfand2004} restricted to about 2 dimensions.

Our method is mainly limited by taking the cholesky decomposition of the block diagonal $K_B$, a 
$ND\nu \times ND\nu$ matrix.  However, 
$\text{chol}(\text{blkdiag}(A,B,\dots) )= \text{blkdiag}(\text{chol}(A),\text{chol}(B),\dots)$. So
in the case with equal length-scales for each dimension, we only need to take the cholesky of an
$N \times N$ matrix $K$, an $\mathcal{O}(N^3)$ operation, independent of dimension!  In the more
general case with $D$ different length-scales, it is an $\mathcal{O}(DN^3)$ operation.  Taking into
account the likelihood, and other operations, the total training complexity is either $\mathcal{O}(N^3 + \nu D^2)$
for equal length-scales, or $\mathcal{O}(DN^3 + \nu D^2)$ using separate length-scales for each dimension.
In the latter more general case, we could in principle go to about 1000 dimensions, assuming $\nu$ is $\mathcal{O}(D)$,
and for instance, a couple years worth of financial training data, which is typical for GARCH \citep{engle2009}.
Using sparse GP techniques we could go to even higher dimensions.  In practice, MCMC may be infeasible for very high $D$, but we have found Elliptical Slice Sampling incredibly robust: we will
shortly include a 200 dimensional experiment in an updated version.  Overall, this is an impressive 
scaling -- without making further assumptions in our model, we can go well beyond 5 dimensions with full generality.

\section{Multivariate GARCH}
We compare predictions of $\Sigma(t)$ made by the generalised Wishart process to those made by
multivariate GARCH (MGARCH), since GARCH \citep{bollerslev86} is extremely popular and 
arguably unsurpassed at predicting the volatility of returns on equity indices and 
currency exchanges \citep{granger05, hansen05, engle2009}.

Consider a zero mean $D$ dimensional vector stochastic process $\bm{x}(t)$ with a 
time changing covariance matrix $\Sigma(t)$ as in \eqref{eqn: xsigma}.  In the general MGARCH framework,
\begin{equation}
\bm{x}(t) = \Sigma(t)^{1/2}\bm{\eta}(t)\,,
\end{equation}
where $\bm{\eta}(t)$ is an i.i.d. vector white noise process with $\mathbb{E}[\bm{\eta}(t) \bm{\eta}(t)^{\top}] = I$,
and $\Sigma(t)$ is the covariance matrix of $\bm{x}(t)$ conditioned on all information up until time $t-1$.

The first and most general MGARCH model, the VEC model of \citet{bollerslev88}, specifies $\Sigma_t$ as
\begin{equation}
 \text{vech}(\Sigma_t) = \bm{a}_0 + \sum_{i=1}^{q} A_i \text{vech}(\bm{x}_{t-i}\bm{x}_{t-i} ^{\top}) + \sum_{j=1}^{p} B_j \text{vech}(\Sigma_{t-j})\,.   
\end{equation}
$A_i$ and $B_j$ are $D(D+1)/2 \times D(D+1)/2$ matrices of parameters, and $\bm{a}_0$ is a $D(D+1)/2 \times 1$ vector of parameters.\footnote{We use $\Sigma(t)$ and $\Sigma_t$ interchangeably.}  The 
vech operator stacks the columns of the lower triangular part of a $D \times D$ matrix into a vector of length $D(D+1)/2$.  For example, 
$\text{vech}(\Sigma) = (\Sigma_{11}, \Sigma_{21},  \dots, \Sigma_{D1}, \Sigma_{22}, \dots, \Sigma_{D2}, \dots, \Sigma_{DD})^{\top}$.  This
model is general, but difficult to use.  There are $(p+q)(D(D+1)/2)^2 + D(D+1)/2$ parameters!  These parameters are hard to interpret,
and there are no conditions under which $\Sigma_t$ is positive definite for all $t$.  \citet{gourieroux97} discusses the challenging (and sometimes
impossible) problem of keeping $\Sigma_t$ positive definite.  Training is done by a constrained maximum likelihood, where the log likelihood is given by
\begin{equation}  
 \mathcal{L} = -\frac{ND}{2}\log(2\pi) - \frac{1}{2} \sum_{t=1}^{N} [\log|\Sigma_t| + \bm{x}_t^{\top}\Sigma_t^{-1}\bm{x}_t]\,,  
\end{equation}
supposing that $\bm{\eta}_t \sim \mathcal{N}(0,I)$, and that there are $N$ training points.

Subsequent efforts have led to simpler but less general models.  We can let $A_j$ and $B_j$ be diagonal matrices.  This model has 
notably fewer (though still $(p+q+1)D(D+1)/2$) parameters, and there are conditions under which $\Sigma_t$ is positive definite
for all $t$ \citep{engle94}.  But now there are no interactions between the different conditional variances and covariances.  A 
popular variant assumes \textit{constant} correlations between the $D$ components of $\bm{x}$, and only lets the marginal variances
-- the diagonal entries of $\Sigma(t)$ -- vary \citep{bollerslev90}.  

We compare to the `full' BEKK variant of \citet{engle1995}, as implemented by Kevin Shepphard in the UCSD GARCH Toolbox.\footnote{http:\///www.kevinsheppard.com\//wiki\//UCSD\_GARCH}
We chose BEKK because it is the most general MGARCH variant in widespread use.  We use the first order model:
\begin{equation}
\Sigma_t = CC^{\top} + A^{\top} \bm{x}_{t-1} \bm{x}_{t-1}^{\top} A + B^{\top} \Sigma_{t-1} B \,,
\end{equation}
where $A, B$ and $C$ are $D \times D$ matrices of parameters.  $C$ is lower triangular to ensure that $\Sigma_t$ 
is positive definite during maximum likelihood training.  
For a full review of multivariate GARCH models, see \cite{anna09}.

\section{Experiments}

In our experiments, we predict the covariance matrix for multivariate observations $\bm{x}(t)$ as 
$\hat{\Sigma}(t) = \mathbb{E}[\Sigma(t)|\mathcal{D}]$.  These experiments closely follow \citep{engle2009},
a rigorous empirical comparison of different GARCH models.  We use a Gaussian likelihood, as 
in \eqref{eqn: likeliwish}, except with a zero mean function.  We make historical predictions, 
and one step ahead forecasts.  Historical predictions are made at observed time points, or between these points.  
The one step ahead forecasts are predictions of $\Sigma(t+1)$ taking into account all observations until time
$t$.  Historical predictions are important -- for example, they can be used to understand
the nature of covariances between different equity indices during a past financial crisis.

To make these predictions we learn distributions over the GWP parameters through 
the Gibbs sampling procedure outlined in section 5. 
The kernel functions we use are solely parametrized by a one dimensional length-scale 
$l$, which indicates how dependent $\Sigma(t)$ and $\Sigma(t+a)$ are 
on one another. We place a lognormal prior on the length-scale, and sample from the posterior 
with axis-aligned slice sampling.  

For each experiment, we choose a kernel function we want to use with the GWP.  We then 
compare to a GWP that uses an Ornstein-Uhlenbeck (OU) kernel function, $k(t,t') = \exp(-|t-t'|/l)$.
Even though we are still taking advantage of the inference procedures in
the GWP formulation, we refer to this variant of GWP as a simple
Wishart process (WP), since the classic \citet{bru91} construction
is like a special case of our generalised Wishart process restricted
to using a one dimensional Gaussian process with an OU covariance
structure.

To assess predictions we use the Mean Squared Error (MSE) between the predicted and
true covariance matrices, which is always safe since we never observe the true 
$\Sigma(t)$. When the truth is not known, we use the proxy $S_{ij}(t) = x_i(t)x_j(t)$,
to harmonize with the econometrics literature.  $x_i$ is the $i^{\text{th}}$ component of the multivariate observation $\bm{x}(t)$. 
This is intuitive because $\mathbb{E}[x_i(t)x_j(t)] = \Sigma_{ij}(t)$, assuming $\bm{x}(t)$ 
has a zero mean.  In a thorough empirical study, \citet{engle2009} use the 
univariate analogue of this proxy.  We do not use likelihood for assessing historical predictions,
since that is a training error (for MGARCH), but we do use log likelihood ($\mathcal{L}$) 
for forecasts.  Although we give results when a proxy is used for historical assessments, 
this is also a sort of training error, and should be observed with caution.

\begin{figure}
\centering
\includegraphics[scale=.67]{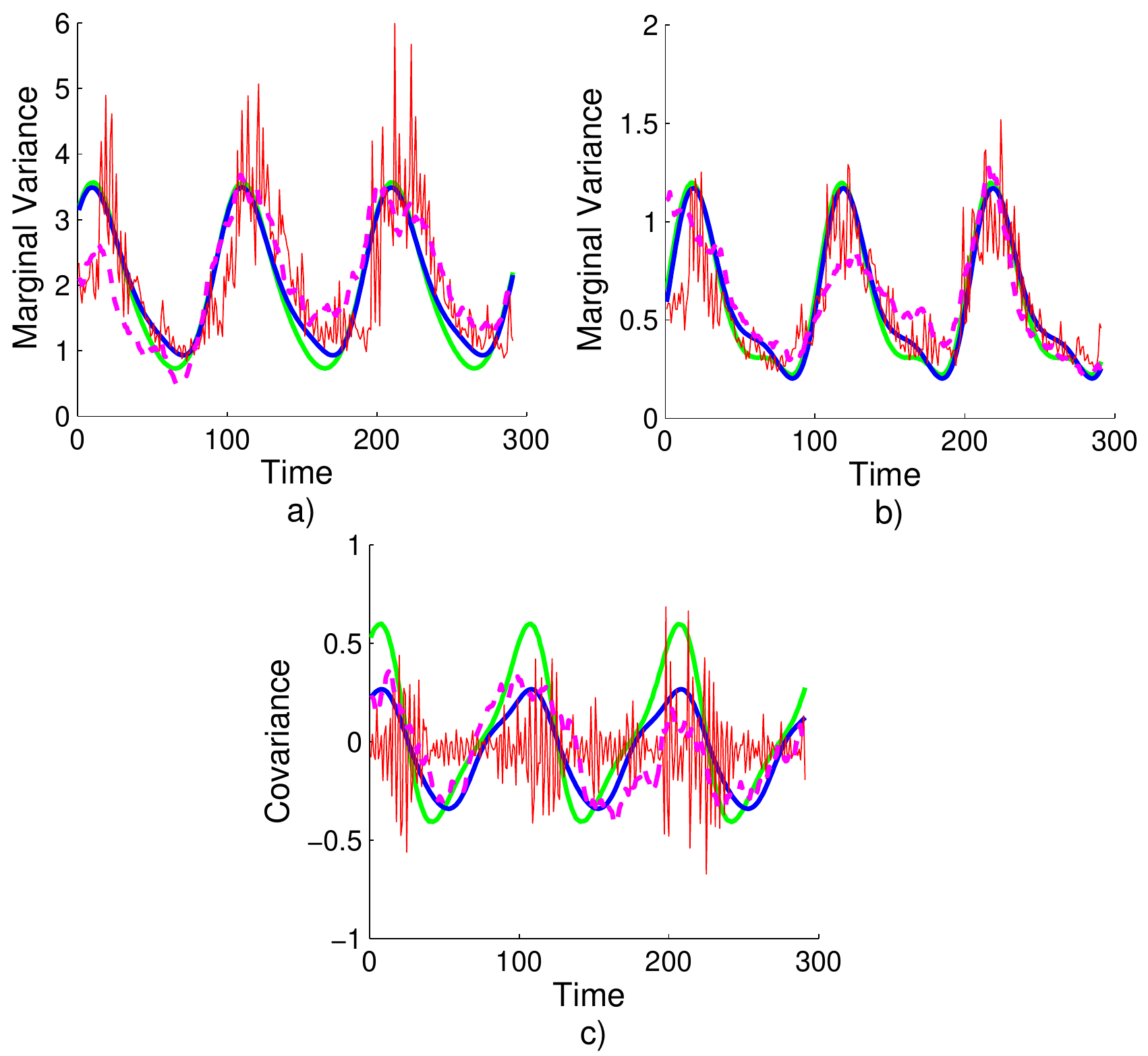}
\caption{\small Reconstructing the historical $\Sigma_p(t)$ for the \texttt{PERIODIC} data set.
We show the truth (green), and GWP (blue), WP (dashed magenta), and MGARCH (thin red) predictions.
a) and b) are the marginal variances (diagonal elements of $\Sigma_p(t)$), 
and c) is the covariance (off-diagonal element of $\Sigma_p(t)$).}
\label{fig: recon}
\end{figure}

We begin by generating a $2 \times 2$ time varying covariance matrix $\Sigma_p(t)$ with 
periodic components, and simulating data at 291 time steps from a Gaussian distribution:
\begin{equation}
 \bm{x}(t) \sim \mathcal{N}(\bm{0},\Sigma_p(t)) \,.
\end{equation}
Periodicity is especially common to financial and climate data, where daily trends 
repeat themselves.  For example, the intraday volatility on equity indices and
currency exchanges has a periodic covariance structure.  \citet{andersen97} 
discuss the lack of -- and critical need for -- models that account for this
periodicity.  In the GWP formulation, we can easily account for this by using
a periodic kernel function.  We reconstruct $\Sigma_p(t)$ using the kernel
$k(t,t') = \exp(-2\sin((t-t')^2)/l^2)$.  We reconstructed 
the historical $\Sigma_p$ at all 291 data points, and after having learned the
parameters for each of the models from the first 200 data points, made one step
forecasts for the last 91 points.  Table \ref{tab: predictions} and Figure \ref{fig: recon} 
show the results.  We call this data set \texttt{PERIODIC}.  The GWP outperforms the competition on 
all error measures. It identifies the periodicity and underlying smoothness of $\Sigma_p$
that neither the WP nor MGARCH accurately discern: both
are too erratic.  And MGARCH is especially poor at learning the time changing
covariance (off-diagonal entry of $\Sigma_p$) in this data set.
\begin{table}
\caption{Error for predicting multivariate volatility.}
\begin{center}
\begin{tabular}{l r r r r r}
\toprule
  & MSE Historical & MSE Forecast & $\mathcal{L}$ Forecast \\
\midrule             
 \texttt{PERIODIC} (2D): &   &   &  \\  
  GWP & $\bm{0.0841}$ & $\bm{0.118}$ & $\bm{-257}$ \\
  WP & $0.458$ & $3.04$ & $-286$ \\
  MGARCH & $0.913$ & $1.95$ & $-270$ \\
\midrule
  \texttt{EXCHANGE} (3D): &  &  &  \\   
  GWP & $\bm{3.49 \times 10^{-8}}$ & $\bm{4.32 \times 10^{-8}}$ & $2020$   \\
  WP & $\bm{3.49 \times 10^{-8}}$ & $6.28 \times 10^{-8}$ & $1950$  \\
  MGARCH & $3.56 \times 10^{-8}$ & $4.45 \times 10^{-8}$ & $\bm{2050}$  \\
\midrule
 \texttt{EQUITY} (5D): &  &  &  \\  
  GWP & $\bm{7.01 \times 10^{-8}}$ & $\bm{1.46 \times 10^{-7}}$ & $\bm{2930}$  \\
  WP & $9.89 \times 10^{-8}$ & $2.23 \times 10^{-7}$ & $1710$  \\
  MGARCH & $16.7 \times 10^{-8}$ & $7.34 \times 10^{-7}$ & $2760$ \\
\bottomrule
\end{tabular}
\end{center}
\label{tab: predictions}
\end{table}

For our next experiment, we predict $\Sigma(t)$ for the returns on 
three currency exchanges -- the Canadian to US Dollar, 
the Euro to US Dollar, and the US Dollar to the Great Britain Pound -- in the period 15/7/2008-15/2/2010; 
this encompasses the recent financial crisis and so is of particular 
interest to economists.  We call this data set \texttt{EXCHANGE}.  
We use the proxy $S_{ij}(t) = x_i(t)x_j(t)$.  With the GWP, we use the squared exponential 
kernel $k(t,t') = \exp(-0.5(t-t')^2/l^2)$.  We make 200 one step ahead forecasts, 
having learned the parameters for each of the models on the previous 200 data points.  
We also make 200 historical predictions for the same data points as the forecasts.
The results are in Table \ref{tab: predictions}. 
\begin{figure}
\centering
\includegraphics[scale=.51]{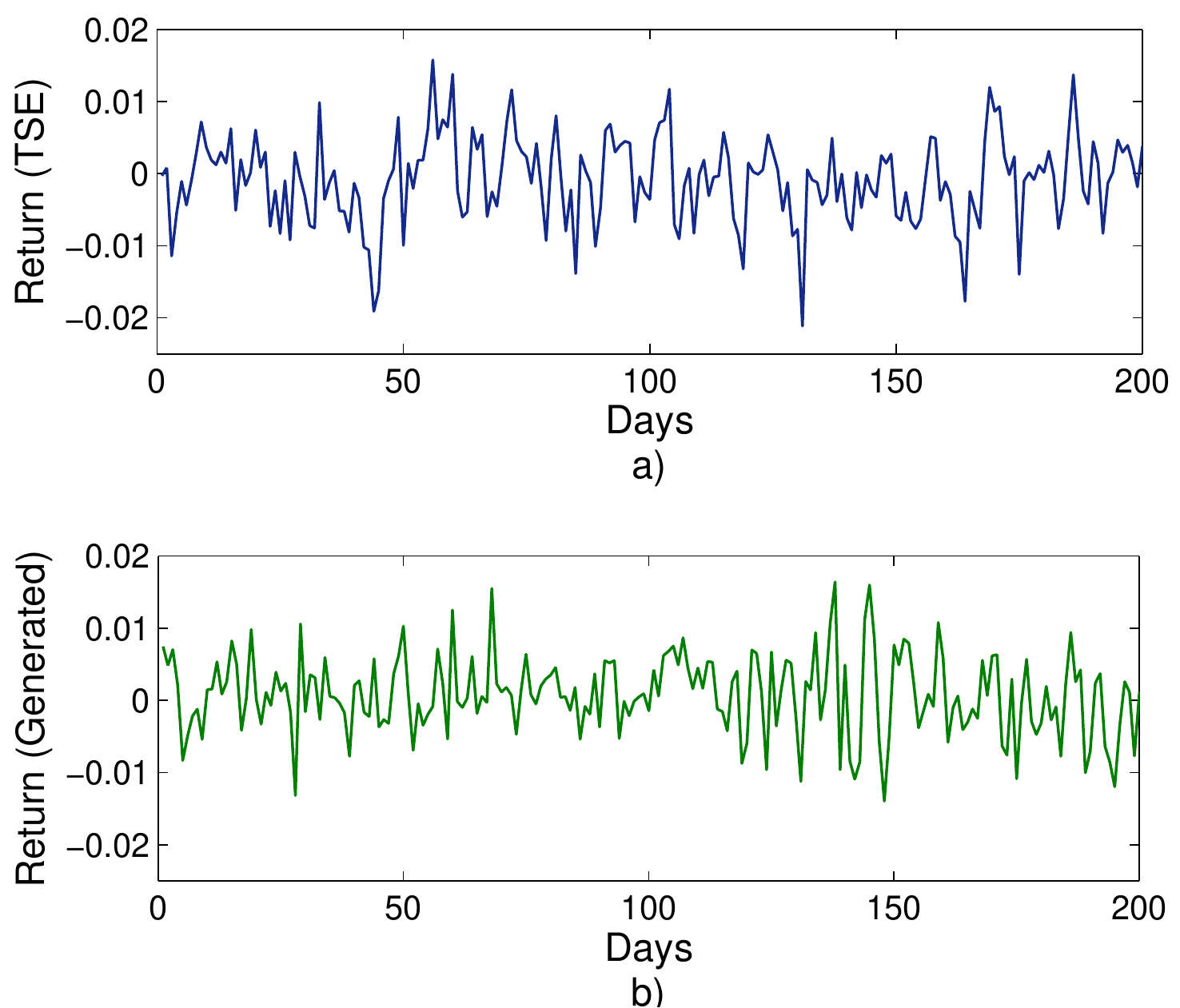}
\caption{\small Generating a return series using the empirical covariance of financial data. 
In a) are daily returns for the Toronto Stock Exchange (TSE).  In b) is 
one component of observations generated from a known time-varying covariance matrix 
which emulates the covariance matrix for returns on equity indices.  These two plots are not supposed to represent 
the same returns (e.g. we could have chosen NASDAQ instead of the TSE and kept panel b) the same), 
so the same trends will not be present.  However, the simulated data has the same type of covariance structure 
as the financial data: it looks like financial data.}
\label{fig: simulated}
\end{figure}

Unfortunately, we cannot properly assess predictions on natural data, because
we do not know the true $\Sigma(t)$.  For example, whether we use MSE with a proxy, 
or likelihood, historical predictions would be assessed with the same data
used for training.  In consideration of this problem, we generated a 
time varying covariance matrix $\tilde{\Sigma}(t)$ based on the 
empirical time varying covariance of the daily returns on five equity 
indices -- NASDAQ, FTSE, TSE, NIKKEI, and the Dow Jones Composite -- over 
the period from 15/2/1990-15/2/2010.\footnote{We define a \textit{return} as 
$r_t = \log(P_{t+1}/P_t)$, where $P_t$ is the price on day $t$.}  We then generated a return series by 
sampling from a multivariate Gaussian at each time step using $\tilde{\Sigma}(t)$.
As seen in Figure \ref{fig: simulated}, the generated return series behaves like 
equity index returns.  This method is not faultless; for example, we assume that the 
returns are normally distributed.  However, the models we compare between also make 
this assumption, and so no model is given an unfair advantage over another.  
And there is a critical benefit: we can compare predictions with the true underlying $\tilde{\Sigma}(t)$.

To make forecasts and historical predictions on this data set (\texttt{EQUITY}), 
we used a GWP with a squared exponential kernel, $k(t,t') = \exp(-0.5(t-t')^2/l^2)$.
We follow the same procedure as before and make 200 forecasts
and historical predictions; results are in Table \ref{tab: predictions}.  
 
Both the \texttt{EXCHANGE} and \texttt{EQUITY} data sets are especially suited to GARCH
\citep{granger05, hansen05, engle2009, bollerslev96, mccullough98, brooks2001}.  However,
the generalised Wishart process outperforms GARCH on both of these data sets.  Based on 
our experiments, there is evidence that the GWP is particularly good at capturing the
co-variances (off-diagonal elements of $\Sigma(t)$) as compared to GARCH.  The GWP also
outperforms the WP, which has a fixed Ornstein-Uhlenbeck covariance structure, even 
though in our experiments the WP takes advantage of the new inference procedures we 
have derived.  Thus the difference in performance is likely because the GWP is capable
of capturing complex interdependencies in volatility, whereas the WP is not.  

\section{Discussion}
We introduced a stochastic process -- the generalised Wishart process (GWP) -- which we used to model
time-varying covariance matrices $\Sigma(t)$.  Unlike the alternatives, the GWP can easily model a diverse
class of covariance structures for $\Sigma(t)$.  In the future, the GWP could be applied to study how these
covariance matrices depend on other variables, like interest rates, in addition to time.  Future research could 
also apply the GWP to extremely high dimensional problems.  

We hope to unify efforts in machine learning and econometrics to inspire
new multivariate volatility models that are simultaneously general, easy to 
interpret, and tractable in high dimensions.

\subsubsection*{Acknowledgements}

Thanks to Carl Edward Rasmussen and John Patrick Cunningham for helpful discussions.  AGW is supported by an NSERC grant.


\bibliographystyle{apalike} 
\bibliography{mbib}

\begin{thebibliography}{}

\bibitem[Andersen and Bollerslev, 1997]{andersen97}
Andersen, T.~G. and Bollerslev, T. (1997).
\newblock Intraday periodicity and volatility persistence in financial markets.
\newblock {\em Journal of Empirical Finance}, 4(2-3):115--158.

\bibitem[Bae et~al., 2003]{bae2003}
Bae, K., Karolyi, G., and Stulz, R. (2003).
\newblock {A new approach to measuring financial contagion}.
\newblock {\em Review of Financial Studies}, 16(3):717.

\bibitem[Bollerslev, 1986]{bollerslev86}
Bollerslev, T. (1986).
\newblock Generalized autoregressive conditional heteroskedasticity.
\newblock {\em Journal of Econometrics}, 31(3):307--327.

\bibitem[Bollerslev, 1990]{bollerslev90}
Bollerslev, T. (1990).
\newblock Modeling the coherence in short-term nominal exchange rates: A
  multivariate generalized arch approach.
\newblock {\em Review of Economics and Statistics}, 72:498--505.

\bibitem[Bollerslev et~al., 1988]{bollerslev88}
Bollerslev, T., Engle, R.~F., and Wooldridge, J.~M. (1988).
\newblock A capital asset pricing model with time-varying covariances.
\newblock {\em The Journal of Political Economy}, 96(1):116--131.

\bibitem[Bollerslev and Ghysels, 1996]{bollerslev96}
Bollerslev, T. and Ghysels, E. (1996).
\newblock Periodic autoregressive conditional heteroscedasticity.
\newblock {\em Journal of Business and Economic Statistics}, 14:139--151.

\bibitem[Bonilla et~al., 2008]{bonilla2008}
Bonilla, E., Chai, K., and Williams, C. (2008).
\newblock {Multi-task Gaussian process prediction}.
\newblock In {\em NIPS}.

\bibitem[Boyle and Frean, 2004]{boyle2004}
Boyle, P. and Frean, M. (2004).
\newblock {Dependent {G}aussian processes}.
\newblock In {\em NIPS}.

\bibitem[Brooks et~al., 2001]{brooks2001}
Brooks, C., Burke, S., and Persand, G. (2001).
\newblock Benchmarks and the accuracy of {{GARCH}} model estimation.
\newblock {\em International Journal of Forecasting}, 17:45--56.

\bibitem[Brownlees et~al., 2009]{engle2009}
Brownlees, C.~T., Engle, R.~F., and Kelly, B.~T. (2009).
\newblock A practical guide to volatility forecasting through calm and storm.
\newblock Available at SSRN: http://ssrn.com/abstract=1502915.

\bibitem[Bru, 1991]{bru91}
Bru, M. (1991).
\newblock Wishart processes.
\newblock {\em Journal of Theoretical Probability}, 4(4):725--751.

\bibitem[Engle and Kroner, 1995]{engle1995}
Engle, R. and Kroner, K. (1995).
\newblock {Multivariate simultaneous generalized ARCH}.
\newblock {\em Econometric theory}, 11(01):122--150.

\bibitem[Engle et~al., 1994]{engle94}
Engle, R., Nelson, D., and Bollerslev, T. (1994).
\newblock Arch models.
\newblock {\em Handbook of Econometrics}, 4:2959--3038.

\bibitem[Engle, 1982]{engle82}
Engle, R.~F. (1982).
\newblock Autoregressive conditional heteroscedasticity with estimates of the
  variance of united kingdom inflation.
\newblock {\em Econometrica}, 50(4):987--1007.

\bibitem[Gelfand et~al., 2004]{gelfand2004}
Gelfand, A., Schmidt, A., Banerjee, S., and Sirmans, C. (2004).
\newblock {Nonstationary multivariate process modeling through spatially
  varying coregionalization}.
\newblock {\em Test}, 13(2):263--312.

\bibitem[Geman and Geman, 1984]{geman1984}
Geman, S. and Geman, D. (1984).
\newblock {Stochastic relaxation, {G}ibbs distributions, and the bayesian
  restoration of images}.
\newblock {\em IEEE Transactions on Pattern Analysis and Machine Intelligence},
  6(2):721--741.

\bibitem[Gouri{\'e}roux, 1997]{gourieroux97}
Gouri{\'e}roux, C. (1997).
\newblock {\em {ARCH models and financial applications}}.
\newblock Springer Verlag.

\bibitem[Gouri{\'e}roux et~al., 2009]{gourieroux09}
Gouri{\'e}roux, C., Jasiak, J., and Sufana, R. (2009).
\newblock {The Wishart autoregressive process of multivariate stochastic
  volatility}.
\newblock {\em Journal of Econometrics}, 150(2):167--181.

\bibitem[Green, 1995]{green1995}
Green, P. (1995).
\newblock {Reversible jump Markov chain Monte Carlo computation and Bayesian
  model determination}.
\newblock {\em Biometrika}, 82(4):711.

\bibitem[Hansen and Lunde, 2005]{hansen05}
Hansen, P.~R. and Lunde, A. (2005).
\newblock A forecast comparison of volatility models: Does anything beat a
  {{GARCH}}(1,1).
\newblock {\em Journal of Applied Econometrics}, 20(7):873--889.

\bibitem[Harvey et~al., 1994]{harvey94}
Harvey, A., Ruiz, E., and Shephard, N. (1994).
\newblock Multivariate stochastic variance models.
\newblock {\em The Review of Economic Studies}, 61(2):247--264.

\bibitem[Markowitz, 1952]{markowitz1952}
Markowitz, H. (1952).
\newblock {Portfolio selection}.
\newblock {\em Journal of Finance}, 7(1):77--91.

\bibitem[McCullough and Renfro, 1998]{mccullough98}
McCullough, B. and Renfro, C. (1998).
\newblock Benchmarks and software standards: A case study of {{GARCH}}
  procedures.
\newblock {\em Journal of Economic and Social Measurement}, 25:59--71.

\bibitem[Merton, 1972]{merton1972}
Merton, R. (1972).
\newblock {An analytic derivation of the efficient portfolio frontier}.
\newblock {\em Journal of financial and quantitative analysis},
  7(4):1851--1872.

\bibitem[Murray et~al., 2010]{murray10}
Murray, I., Adams, R.~P., and MacKay, D.~J. (2010).
\newblock Elliptical {S}lice {S}ampling.
\newblock {\em JMLR: W\&CP}, 9:541--548.

\bibitem[Philipov and Glickman, 2006]{philipov06}
Philipov, A. and Glickman, M. (2006).
\newblock {Multivariate stochastic volatility via Wishart processes}.
\newblock {\em Journal of Business and Economic Statistics}, 24(3):313--328.

\bibitem[Poon and Granger, 2005]{granger05}
Poon, S.-H. and Granger, C.~W. (2005).
\newblock Practical issues in forecasting volatility.
\newblock {\em Financial Analysts Journal}, 61(1):45--56.

\bibitem[Rasmussen and Williams, 2006]{rasmussen06}
Rasmussen, C.~E. and Williams, C.~K. (2006).
\newblock {\em Gaussian processes for Machine Learning}.
\newblock The {MIT} {P}ress.

\bibitem[Robert and Casella, 2004]{robert2004}
Robert, C. and Casella, G. (2004).
\newblock {\em {Monte Carlo statistical methods}}.
\newblock Springer Verlag.

\bibitem[Sharpe, 1966]{sharpe1966}
Sharpe, W. (1966).
\newblock {Mutual fund performance}.
\newblock {\em Journal of business}, 39(1):119--138.

\bibitem[Silvennoinen and Ter{\"a}svirta, 2009]{anna09}
Silvennoinen, A. and Ter{\"a}svirta, T. (2009).
\newblock Multivariate {GARCH} models.
\newblock {\em Handbook of Financial Time Series}, pages 201--229.

\bibitem[Teh et~al., 2005]{teh2005}
Teh, Y., Seeger, M., and Jordan, M. (2005).
\newblock {Semiparametric latent factor models}.
\newblock In {\em Workshop on Artificial Intelligence and Statistics},
  volume~10.

\bibitem[Uhlenback and Ornstein, 1930]{uhlenbeck30}
Uhlenback, G. and Ornstein, L. (1930).
\newblock On the theory of brownian motion.
\newblock {\em Phys. Rev.}, 36:823--841.

\bibitem[Wilson and Ghahramani, 2010]{wilson2010}
Wilson, A. and Ghahramani, Z. (2010).
\newblock Copula processes.
\newblock In Lafferty, J., Williams, C. K.~I., Shawe-Taylor, J., Zemel, R., and
  Culotta, A., editors, {\em Advances in Neural Information Processing Systems
  23}, pages 2460--2468.

\end{thebibliography}

\end{document}